# Formation of Black Holes from Collapsed Cosmic String Loops

R. R. Caldwell

*University of Cambridge, D.A.M.T.P.*
*Silver Street, Cambridge CB3 9EW, U.K.*
*email: R.R.Caldwell@amtp.cam.ac.uk*

Paul Casper

*Department of Physics, University of Wisconsin – Milwaukee*
*P.O. Box 413, Milwaukee, Wisconsin 53201, U.S.A.*
*email: pcasper@dirac.phys.uwm.edu*

(August 25, 1995)

## Abstract

The fraction of cosmic string loops which collapse to form black holes is estimated using a set of realistic loops generated by loop fragmentation. The smallest radius sphere into which each cosmic string loop may fit is obtained by monitoring the loop through one period of oscillation. For a loop with invariant length $L$ which contracts to within a sphere of radius $R$, the minimum mass-per-unit length $\mu_{\min}$ necessary for the cosmic string loop to form a black hole according to the hoop conjecture is $\mu_{\min} = R/(2GL)$. Analyzing $25,576$ loops, we obtain the empirical estimate $f_{\rm BH} = 10^{4.9\pm0.2}(G\mu)^{4.1\pm0.1}$ for the fraction of cosmic string loops which collapse to form black holes as a function of the mass-per-unit length $\mu$ in the range $10^{-3} \lesssim G\mu \lesssim 3 \times 10^{-2}$. We use this power law to extrapolate to $G\mu \sim 10^{-6}$, obtaining the fraction $f_{\rm BH}$ of physically interesting cosmic string loops which collapse to form black holes within one oscillation period of formation. Comparing this fraction with the observational bounds on a population of evaporating black holes, we obtain the limit $G\mu \leq 3.1(\pm 0.7) \times 10^{-6}$ on the cosmic string mass-per-unit-length. This limit is consistent with all other observational bounds.

Typeset using REVTEX

# I. INTRODUCTION

Cosmic strings are topologically stable objects which may have formed as a result of a phase transition in the early universe. Strings which have a mass-per-unit length, $\mu$, such that $G\mu \sim 10^{-6}$, where $G$ is Newton's constant, may play a significant role in the formation of the large scale structures observed in the Universe today. (See the monograph by Vilenkin & Shellard [1] for a comprehensive review of the physics of cosmic strings.) Cosmic strings may display a number of other observable features through which the one free parameter of the cosmic string scenario, $\mu$, may be constrained. The nucleosynthesis limit on relativistic particle species, and the pulsar timing limit on a stochastic gravitational wave background provide observational constraints on the spectrum of gravitational waves emitted by cosmic strings, and thus on $\mu$. Another observational constraint comes from the anisotropies induced in the cosmic microwave background by a network of cosmic strings. An upper bound of

$$G\mu \lesssim 1 - 4 \times 10^{-6} \tag{1.1}$$

is required for cosmic strings to be compatible with these observational constraints [2,3].

In this paper, we evaluate an additional observational constraint on the mass-per-unit length of cosmic strings. This restriction is due to observational constraints on cosmic rays emitted by evaporating primordial black holes which formed from collapsed cosmic string loops. Previous attempts have been made to evaluate this constraint, and to estimate the fraction of cosmic string loops which collapse to form black holes [4–6]. In an earlier paper by Caldwell & Gates [7], observational restrictions on the fraction of cosmic string loops which collapse to form black holes were derived using an updated model for the evolution of a cosmic string network. In this paper, *we estimate the fraction of realistic cosmic string loops which collapse to form black holes as a function of their mass-per-unit length.* Combining this function with the above observational restriction gives an additional constraint on the mass-per-unit length of cosmic strings.

The organization of the paper is as follows. In section II we describe the evolution by which a cosmic string loop may collapse to form a black hole. In section III we describe the numerical methods used to generate realistic cosmic string loop trajectories [8], and to determine the minimum mass-per-unit length necessary for a given loop to form a black hole. In section IV we present the results of our numerical analysis, obtaining the fraction $f_{\rm BH}$ of cosmic string loops which collapse to form black holes as a function of $\mu$. We compare this fraction with observational bounds in section V, obtaining a bound on $\mu$. We conclude in section VI.

Throughout this paper we use units in which the speed of light $c = 1$.

# II. COSMIC STRING EVOLUTION

In this section we begin with a review of the cosmic string equations of motion. We then describe the conditions under which a cosmic string loop may collapse to form a black hole.

In flat space-time, the equations of motion for the cosmic string trajectory $\mathbf{x}(\sigma, \tau)$ are

$$\ddot{\mathbf{x}} = \mathbf{x}'' \qquad \text{with gauge conditions} \qquad \dot{\mathbf{x}} \cdot \mathbf{x}' = 0 \qquad \dot{\mathbf{x}}^2 + \mathbf{x}'^2 = 1, \tag{2.1}$$

where $\tau$ is the proper time in the string center-of-mass frame, $\sigma$ is a space-like parameter which runs along the string, $\cdot \equiv \partial/\partial\tau$, and $' \equiv \partial/\partial\sigma$. The trajectory for a closed cosmic string loop in the center-of-mass frame is periodic in both $\tau$ and $\sigma$:

$$\mathbf{x}(\sigma,\tau) = \mathbf{x}(\sigma \pm L,\tau) = \mathbf{x}(\sigma,\tau \pm L). \tag{2.2}$$

Here, $L$ is the total invariant length of the string loop. The rest energy and angular momentum of the loop are given by

$$M = \mu \int_0^L d\sigma = \mu L, \qquad \mathbf{J} = \mu \int_0^L d\sigma \, \mathbf{x}(\sigma,\tau) \times \dot{\mathbf{x}}(\sigma,\tau). \tag{2.3}$$

Thus for a given loop with a trajectory $\mathbf{x}(\sigma,\tau)$ and invariant length $L$ we may obtain the loop rest mass and angular momentum.

To determine whether a cosmic string loop collapses to form a black hole we assume that the hoop conjecture [9] is true. According to the hoop conjecture, a black hole with a horizon forms when and only when a mass $M$ gets compacted into a region whose proper circumference $\mathcal{C}$ in every direction is $\mathcal{C} \leq 4\pi GM$ [9]. Hence, within the framework of linearized gravity, a cosmic string loop oscillating in a Minkowski space-time background will collapse to form a black hole when the loop contracts to within a sphere of radius equal to the Schwarzschild radius of its mass. We define the function $R(\tau)$ to be the radius of the smallest sphere circumscribing the cosmic string loop at time $\tau$. We assume that a loop of invariant length $L$ will collapse to form a black hole if at some time $\tau$ the inequality

$$R(\tau) \leq R_S = 2GM = 2G\mu L \tag{2.4}$$

is satisfied.

The collapse of a cosmic string loop to within its Schwarzschild radius is a necessary, but not sufficient condition for the loop to form a black hole. There are several possible barriers to the formation of a black hole, which we discuss next.

### A. Angular Momentum

A rotating loop which collapses to form a black hole will produce a rotating black hole. In order that the curvature singularity within the newly formed rotating black hole of mass $M$ lies within the event horizon, the black hole angular momentum $\mathbf{J}_{BH}$ must satisfy the inequality [1,10]

$$|\mathbf{J}_{BH}| \leq GM^2. \tag{2.5}$$

Hence, the angular momentum of the loop which collapses to form a black hole must also satisfy this inequality. For a particular loop configuration, this bound may be expressed as

$$G\mu \geq L^{-2} \int_0^L d\sigma \, \mathbf{x}(\sigma,\tau) \times \dot{\mathbf{x}}(\sigma,\tau). \tag{2.6}$$

A cosmic string loop will collapse to form a black hole only if its mass-per-unit length, $\mu$, is such that both (2.4) and (2.6) are satisfied.

## B. Radiation

As a loop collapses it may radiate a sufficient amount of energy such that its minimum configuration never falls within its Schwarzschild radius. Cosmic strings may emit the quanta composing the string, and gravitational radiation. We consider the effects of both types of radiation in order.

Emission of non-gravitational radiation by a cosmic string is suppressed relative to the emission of gravitational radiation unless the curvature along the string is comparable to or smaller than than the loop thickness, $\lambda$, given by the Compton wavelength of the quanta composing the cosmic string [1]. When the radius of curvature along the loop becomes smaller than the string thickness, the string is no longer topologically bound. In such an event, the loop may rapidly "unwind" and vanish in a burst of radiated quanta. Thus we conservatively require that a loop may collapse to form a black hole only if its thickness is much smaller than the Schwarzschild radius

$$\lambda \ll R_{\rm S} = 2GM. \tag{2.7}$$

This inequality has already been taken into account in obtaining the cosmological bound on the fraction of loops which collapse to form black holes [7], and is assumed to hold true for all loops examined in this paper.

Cosmic strings radiate energy in the form of gravitational waves at a rate

$$\dot{E} = \gamma G \mu^2 \tag{2.8}$$

where $\gamma$ is a dimensionless coefficient (typically $\sim 50$), obtained by averaging the radiation power over one period of oscillation. The coefficient $\gamma$ depends upon the specific loop configuration but is independent of the overall loop size for loops well inside the horizon. The product $\gamma G \mu$ characterizes the amplitude of the gravitational back-reaction. Within one period of oscillation, a loop will radiate approximately a fraction $\gamma G \mu / 2$ of its rest mass. In this paper we work in the weak field limit, ignoring the decrease in the loop's rest mass and the effects of gravitational back-reaction which occur during a single period oscillation. This is an excellent approximation for GUT-scale cosmic string loops, where $G\mu \lesssim 10^{-6}$. While the function $f_{\rm BH}(\mu)$ found in section IV is defined over a large range in $\mu$, it may only be consistently used in the region where $\mu$ is small enough for the weak field approximation to be valid. The only predictions made using the function $f_{\rm BH}(\mu)$ are for cosmologically interesting loops with $G\mu \sim 10^{-6}$.

## III. NUMERICAL METHODS

To develop an empirical estimate of the fraction of cosmic string loops which collapse to form black holes, we examine a large set of realistic string loops generated by loop fragmentation. In this section we describe the numerical methods used to generate these loops and to determine the minimum mass-per-unit length required for each loop to satisfy the hoop conjecture.

## A. Generation of Loop Trajectories

The cosmic string loop trajectories examined in this paper were generated using the numerical loop fragmentation code developed by Casper and Allen [8] based on earlier work by Scherrer and Press [11]. This code takes initial (parent) loops which are much shorter than the cosmological Hubble length and evolves them forward in time. Because the evolution is only followed for a short time (through a single period of oscillation or less) the effect of the gravitational back-reaction on the loop trajectories is neglected. The loops are evolved forward until they either self-intersect or complete one full oscillation without self-intersecting. If a loop of cosmic string self-intersects, it will fragment into two child string loops. The trajectories of the child loops can be determined analytically in terms of the known initial parent loop trajectory [11]. Thus, the trajectory of each child loop can be calculated with the same accuracy as that of its initial parent loop. Once the parent loop self-intersects, the fragmentation code recursively evolves and fragments each child loop until only non-self-intersecting loops remain.

A realistic set of string loop trajectories may be generated by loop fragmentation in the following way [11]. Parent loops with very different initial conditions are recursively fragmented into non-self-intersecting child loops. If the properties of the final child loops are similar, independent of the initial conditions used for their parent loops, then they should be representative of the properties of realistic cosmic string loops. The cosmic string loops examined in this paper are the non-self-intersecting child loops resulting from the fragmentation of a large number of parent loops having four very different sets of initial conditions. We have found that, despite the differences in the initial parent loops, the probability distributions for the formation of black holes found for the different sets of child loops are virtually identical.

All of the initial parent loop trajectories (as well as the subsequent child loop trajectories) must be solutions to the equations of motion (2.1). The most general solution to these equations is given by

$$\mathbf{x}(t,\sigma) = \frac{1}{2}[\mathbf{a}(t+\sigma) + \mathbf{b}(t-\sigma)]. \tag{3.1}$$

The pair of functions $\mathbf{a}$ and $\mathbf{b}$ are constrained by the gauge conditions (2.1) to satisfy $|\mathbf{a}'(u)| = |\mathbf{b}'(v)| = 1$, where here the primes denote differentiation with respect to the function's argument. In the center-of-mass frame of the string loop, $\mathbf{a}(u) \equiv \mathbf{a}(u+L)$ and $\mathbf{b}(v) \equiv \mathbf{b}(v+L)$. Because the functions $\mathbf{a}$ and $\mathbf{b}$ are periodic in this frame, each can be described by a closed curve in 3-space. Therefore, we refer to these functions as the $\mathbf{a}$-loop and the $\mathbf{b}$-loop. Together, the $\mathbf{a}$- and $\mathbf{b}$-loops define the trajectory of the string loop. The $\mathbf{a}$- and $\mathbf{b}$-loops defining the initial parent loops fragmented to generate the child loops used in this paper are described next.

The child loops analyzed in this paper are descended from several very different sets of initial parent loops. The $\mathbf{a}$- and $\mathbf{b}$-loops defining the parent loops in each set have the general form

$$a_x(s) = \sum_{m=1}^{\mathcal{M}} a_{xm} \cos(ms + \phi_{xm}), \tag{3.2}$$

with similar equations for $a_y, a_z$, and **b**. (Note that $s$ is not the length along the **a**-loop. The actual length, $u(s)$ must be computed numerically.) For each loop, the $\phi$'s are taken to be random numbers in the range $[0, 2\pi]$. The first set of parent loops (referred to as type A loops), have $\mathcal{M} = 10$ modes and $\mathbf{a}_m$ and $\mathbf{b}_m$ coefficients chosen to be random numbers between 0 and 1. This gives equal amplitude (on average) to both the high and low frequency modes, resulting in convoluted initial loops which each fragment into a large number ($\sim 29$) of child loops. The second set of parent loops (referred to as type B loops) also have $\mathcal{M} = 10$ modes, however for these loops the $\mathbf{a}_m$ and $\mathbf{b}_m$ coefficients are chosen to be random numbers between 0 and $1/m^2$. This gives smaller amplitude to the high frequency modes and results in loops which are less convoluted than the type A loops. As one would expect, the type B loops typically fragment into a much smaller number ($\sim 10$) of child loops. The child loops resulting from fragmenting these first two types of parent loops have been carefully studied in [8]. For the investigation carried out in this paper, the child loops generated by two additional sets of parent loops have also been examined.

To determine whether the properties of the child loops depend upon the number of modes used to define the parent loops, we have fragmented two additional sets of loops. The parent loops from the first of these sets (referred to as type C loops) have only five modes ($\mathcal{M} = 5$) and, like the type B loops, have $\mathbf{a}_m$ and $\mathbf{b}_m$ coefficients chosen to be random numbers between 0 and $1/m^2$. These loops are even less convoluted than the type B loops and only fragment into a small number ($\sim 6$) of child loops. The final set of parent loops (referred to as type D loops) have twenty modes ($\mathcal{M} = 20$) and, like the type A loops, have $\mathbf{a}_m$ and $\mathbf{b}_m$ coefficients chosen to be random numbers between 0 and 1. The type D loops are highly convoluted and fragment into a very large number ($\sim 55$) of child loops.

For the non-self-intersecting child loops to be considered representative of realistic cosmic string loops, their properties must be similar, independent of which initial parent loops they descended from. In reference [8] the non-self-intersecting child loops descended from parent loops of types A and B were examined. It was found that properties of these loops such as the distributions of gravitational radiation rates were independent of the parent loop's initial conditions. In section IV below, we find that the empirical fraction of cosmic string loops which collapse to form black holes as a function of their mass-per-unit length is identical for all four sets of loops. Because the four sets of initial parent loops have such a variety of different initial configurations, we are confident that the empirical fraction found in section IV is a generic property of cosmic string loops formed by fragmentation.

### B. Minimal Bounding Spheres

This section outlines the numerical methods used to determine the minimum radius sphere which completely encloses a cosmic string loop at each instant in time, and the smallest such sphere attained during one period of the loop's oscillation. Once the radius of the smallest bounding sphere attained by a loop is known, it can be compared to the Schwarzschild radius of the loop given as a function of the loop's mass-per-unit length $\mu$. This allows the smallest value of the mass-per-unit length, $\mu_{\min}$, necessary for the loop to satisfy the hoop conjecture to be determined. The distribution of $\mu_{\min}$ values found for all the child loops examined in this paper are used in section IV to find the empirical fraction of cosmic string loops which collapse to form black holes.

The minimum radius sphere completely enclosing a string loop can be found numerically at any time during the loop's oscillation in the following way. (In this section we assume that the loop is in its center-of-mass frame. The child loops generated by loop fragmentation are not typically given in this frame and must first be boosted into it.) Approximate the loop by picking a large number, $N$, of discrete points along the loop's position at the time of interest. Now find the smallest sphere which contains all $N$ points using the algorithm given below. If $N$ is large, this sphere will be an excellent approximation to the sphere containing the loop. For the loops examined in this paper, $N = 20,000$ points were used. The points were evenly spaced in $\sigma$ with additional points added at the exact locations of any kinks on the loop.

The unique smallest sphere containing $N$ points can be found in order $N \log N$ operations. This sphere will be defined by either two, three or four of the $N$ points. The defining points will all lie on the sphere's surface. The remaining points will either lie inside the sphere or on the sphere's surface. If the minimum sphere is defined by two points, then these points will be antipodal points on the sphere, with the line connecting them defining the sphere's diameter. If the minimum sphere is defined by three points, then the unique circle passing through these points defines a great circle on the sphere. If four points are required, then the minimum sphere is simply the unique sphere passing through all four points. General formulae giving the radius and the position of the center of a sphere defined by two, three, or four arbitrary points in 3-space are easily derived.

If the total number of points $N$ is very small, then the minimum sphere enclosing the points can be found by brute force. Simply examine all the spheres defined by each pair, triplet and quadruplet of points. A subset of these spheres will completely enclose all $N$ points, and the smallest sphere in this subset is the one desired. Because this brute force approach requires the examination of order $N^4$ spheres, it is not practical when $N$ is large. However, when $N = 5$, there are only a total of 25 spheres which must be examined. Thus, the minimum sphere enclosing five points can be found very efficiently. This can be exploited to help find the minimum sphere enclosing a large number of points in the following way. First, pick five of the $N$ points at random. Then, using the brute force method, find the minimum sphere enclosing all five of the points. Now compare the positions of all $(N-5)$ remaining points to this sphere (this requires order $N$ operations). If all of the remaining points lie on or inside the sphere then the sphere is the minimum sphere enclosing all $N$ points and the problem is solved. If however one or more of the remaining points lies outside the sphere, then locate the point which lies the furthest from the center of the sphere. Now, the minimum sphere found to enclose the original five points will be defined by either two, three, or four of these five points. Thus, at least one of the original five points was not necessary for defining the minimum sphere. Replace this extraneous point with the point located the furthest from the center of the sphere to give a new set of five points. Now repeat the process by finding the minimum sphere enclosing this new set of five points and comparing the positions of the remaining $(N-5)$ points to this new sphere. Each time one or more points lies outside the sphere, a new iteration must be carried out. The iteration process is guaranteed to terminate because the radius of the minimum sphere enclosing the five points increases with each iteration. The iteration process ends when the minimum sphere enclosing the five points also encloses all of the other $(N-5)$ points. This sphere will be the minimum radius sphere enclosing all $N$ points. We have found that by replacing the extraneous point with the point farthest outside the sphere, order $\log N$ iterations will be required. Thus, the minimum sphere enclosing $N$ points is found with order $N \log N$ operations.

The radius of the minimum sphere enclosing a cosmic string loop changes as a function of time as the loop oscillates. At some time $\tau_{\min}$ during the loop's oscillation, the radius of the bounding sphere will attain an absolute minimum value. Figure 1 shows the radius of the bounding sphere as a function of time over one period of a typical loop's oscillation. The time $\tau_{\min}$ was determined for each of the loops investigated in this paper by a series of approximations. First, the loop's period of oscillation was divided into 25 equal time steps. The minimum bounding sphere was determined at each time step using the method described above. The time which had the smallest bounding sphere (denoted $\tau_1$) was taken as the first approximation to $\tau_{\min}$. This approximation was then improved by dividing a portion of the loop's period (equal to one tenth of the total period) centered on $\tau_1$ into 20 time steps. Determining the minimum bounding sphere at each of these time steps gave a new, more accurate approximation, $\tau_2$, to the exact time $\tau_{\min}$. The approximation to $\tau_{\min}$ was then refined twice more. At each stage, a smaller portion of the loop's period (equal to one tenth the interval examined in the previous stage) was divided into 20 time steps and a more accurate approximation to $\tau_{\min}$ was determined. In this way, we were able to quickly arrive at accurate approximations to $\tau_{\min}$, and thus the radius of the absolute minimum bounding sphere, for each loop. Using this method, the errors in the radii of the absolute minimum bounding spheres were typically less than 0.0001%.

The numerical methods described in this section allow one to determine the radius of the absolute smallest bounding sphere enclosing a string loop as the loop oscillates. Once the radius of this sphere has been calculated, it can be equated with the Schwarzchild radius of the loop to determine the minimum value of the mass-per-unit length, $\mu_{\min}$, required for the loop to satisfy the hoop conjecture. The results of applying these numerical methods to a large number of cosmic string loops are presented in the next section. The distribution of $\mu_{\min}$ values is used to find the empirical fraction of cosmic string loops which collapse to form black holes as a function of their mass-per-unit length.

## IV. RESULTS OF NUMERICAL ANALYSIS

In this section we obtain an empirical estimate of the fraction of cosmic string loops which collapse to form black holes. First, we present the results of the numerical analysis of the 25,576 child loops descended from the four families of parent loops. Next, we compute the effects of the angular momentum barrier to the formation of black holes. Finally, we give a heuristic argument supporting the results of our numerical analysis.

We have analyzed a total of 25,576 loops using the numerical methods outlined in section III. For each loop, the minimum mass-per-unit length, $\mu_{\min}$, required for the loop to satisfy the hoop conjecture was determined by equating the loop's Schwarschild radius with the radius of its absolute minimum bounding sphere (see equation (2.4)). The resulting distribution of $\mu_{\min}$ values was then used to determine the fraction, $f_{\text{BH}}(G\mu)$, of the loops for which $\mu_{\min} \leq \mu$. Thus, $f_{\text{BH}}(G\mu)$ gives the fraction of loops which collapse to form black holes as a function of their mass-per-unit length. Separate empirical fractions are shown in figures 2-5 for the child loops descended from each of the four initial parent loop types (described in section III). Figure 2 shows a $\log - \log$ plot of $f_{\rm BH}(G\mu)$ for the $5,654$ child loops descended from type A parent loops. Similar plots are shown in figures 3-5 for the $6,622$, $11,779$ and $1,521$ child loops descended from parent loops of types B, C and D, respectively. All four plots are very similar, independent of the initial parent loop's type. For $\log_{10}(G\mu) \gtrsim -1.2$, the fraction of loops which form black holes goes to unity in all four figures. This simply shows that within the weak field approximation, all of the child loops will fall inside their Schwarzchild radius as they oscillate if their mass-per-unit length is such that $\log_{10}(G\mu) \gtrsim -1.2$. For $\log_{10}(G\mu) \lesssim -1.2$, all four plots have a linear region with a slope of $\sim 4$. In each case, statistical noise obscures the linear nature of the plots for the lowest values of $\log_{10}(G\mu)$. This noise is due to the small number of loops which are represented by the lower portions of the graphs. The noise is largest in figure 5 where the smallest total number of child loops have been examined.

Because the separate empirical fractions shown in figures 2-5 are very similar, independent of the rather different initial parent loop configurations, we have combined all four sets of results to obtain one cumulative distribution $f_{\rm BH}(G\mu)$ based on all $25,576$ loops. A $\log - \log$ plot of the resulting distribution is shown in figure 6. We have fit this cumulative plot to the linear form $y = ax + b$ over the interval $x \epsilon [-2.0, -1.3]$, where $y = \log_{10} f_{\rm BH}$ and $x = \log_{10}(G\mu)$. From this fit we obtain the functional form of the fraction of loops which collapse to form black holes,

$$f_{\rm BH}(G\mu) = 10^b (G\mu)^a, \quad \text{with} \quad a = 4.0 \pm 0.1 \quad \text{and} \quad b = 4.8 \pm 0.2. \tag{4.1}$$

Thus, $f_{\rm BH}(G\mu)$ is found to be well described by a power law.

We now consider what effect the angular momentum constraint on black hole formation has on the functional form of $f_{\rm BH}(G\mu)$ given in (4.1). The $\mu_{\rm min}$ value found by equating a loop's Schwarzschild radius with the radius of its minimum bounding sphere may not necessarily satisfy the angular momentum constraint (2.6). To satisfy the angular momentum constraint, the value of $\mu_{\rm min}$ found above for each child loop has been examined, and increased as needed until equation (2.6) is satisfied. Following this procedure, approximately 11% of the $\mu_{\rm min}$ values were increased in order to satisfy the angular momentum constraint. The density of the distribution of values of the fractional change in $\mu_{\rm min}$ is a rapidly decreasing function, peaked at zero, with an average fractional change of approximately 30%. The fraction of loops which collapse to form black holes was then recalculated using the modified distribution of $\mu_{\rm min}$ values. The resulting cumulative distribution, for all loop types, is shown in figure 7, along with the earler result from figure 6. The two plots are found to be very similar. A linear fit to the new cumulative distribution shown in figure 7 yields

$$f_{\rm BH}(G\mu) = 10^b (G\mu)^a, \quad \text{with} \quad a = 4.1 \pm 0.1 \quad \text{and} \quad b = 4.9 \pm 0.2. \tag{4.2}$$

Thus, we find that after taking account of the angular momentum constraint, the fraction of loops which collapse to form black holes is still well described by a power law.

We have seen that the functional form given in (4.2) for the fraction, $f_{\rm BH}(G\mu)$, of loops which collapse to form black holes appears to be generic to all cosmic string loops formed by loop fragmentation. To make statements about cosmologically interesting cosmic string loops, we assume that this functional form holds true down to values of $G\mu \sim 10^{-6}$. (Note that making statements about loops which have a mass-per-unit length in this range is consistent with working in the weak field approximation.) Using (4.2), we find the fraction of cosmologically interesting cosmic string loops which collapse to form black holes to be

$$f_{\text{BH}}(10^{-6}) = 2.0(\pm 1.5) \times 10^{-20}. \tag{4.3}$$

In the next section, we will compare this result with the observational restrictions on a population of primordial black holes formed by collapsed cosmic string loops.

Finally, we present a heuristic argument for the functional form found for the fraction of cosmic string loops which collapse to form black holes. For a cosmic string loop to form a black hole, the mass of the loop must contract into a region whose circumference in every direction is less than $4\pi G L \mu$ [9]. That is, the loop must contract to within a fraction of $\sim G\mu$ of its invariant length $L$ in each direction. Assuming that a loop which contracts to within a fraction $\delta$ of its invariant length in any one of the three spatial directions occurs with probability $\delta$, then the probability that a loop will collapse to within a fraction $\delta$ of its invariant length in all three directions is given by

$$P(\delta) \propto \delta^3. \tag{4.4}$$

It then follows that the fraction of loops which collapse in all three directions to within a fraction of their invariant length *smaller* than some value $\Delta$ is given by

$$f(\Delta) = \int_0^\Delta d\delta\, P(\delta) \propto \Delta^4. \tag{4.5}$$

Hence, we would expect the fraction of loops which collapse to form black holes to have the functional form of a power law with $f_{BH}(G\mu) \propto (G\mu)^4$, which is in excellent agreement with our numerical results. If one accepts this heuristic arguement, then the results of this paper may be used to determine the overall normalization of (4.5).

## V. COSMOLOGICAL CONSTRAINTS

The cosmological constraint on the fraction of cosmic string loops which collapse to form black holes was obtained by Caldwell & Gates [7]. The observational restrictions on the cosmic rays emitted by a population of evaporating, primordial black holes [12] were translated into a restriction on the fraction of cosmic string loops which collapse to form black holes, $f_{\text{BH, obs}}$, for loops with a mass-per-unit length $G\mu = 10^{-6}$. It was found that

$$f_{\text{BH, obs}}(G\mu = 10^{-6}) \leq 10^{-17}. \tag{5.1}$$

This bound is due to the present-day abundance of $M \sim 10^{15}$ g black holes emitting photons of energy $\sim 100$ MeV. From (4.2), we find that a fraction $f_{\text{BH}}(G\mu = 10^{-6}) = 2.0(\pm 1.5) \times 10^{-20}$ of cosmic string loops collapse to form black holes. Comparing with (5.1) for the limit on the diffuse $\gamma$−ray background, we see that the predicted fraction is consistent with the maximum value allowed by observations.

The observational bound on the fraction $f_{\text{BH, obs}}$ found in [7] may be evaluated for other values of $G\mu$. The emission of cosmic rays from a population of primordial black holes formed by collapsed cosmic string loops varies as $\mu^{3/2}$, and the upper limit on $f_{\text{BH, obs}}$ is given by

$$f_{\text{BH, obs}} \lesssim 2 \times 10^{-29} (G\mu)^{-2} \left(\frac{\gamma G\mu}{\alpha}\right)^{1/2}. \tag{5.2}$$

In this equation, the dimensionless parameter $\alpha$ characterizes the size of newly formed loops as a fraction of the Hubble radius. While the actual value of $\alpha$ is unknown, it is expected to lie in the range $0.1 \lesssim \alpha/\gamma G\mu \lesssim 1$. The upper limit on $\alpha$ is due to current pulsar timing constraints [3]. The lower limit on $\alpha$, while not firm, is due to the long string gravitational radiation back-reaction, which is generally weaker than the back-reaction on loops. Combining the bound given in equation (5.2) with the predicted fraction (4.2), using $\alpha = \gamma G\mu$ gives an observational bound of $G\mu \leq 3.1(\pm 0.7) \times 10^{-6}$ on the mass-per-unit length of cosmic strings. This bound is consistent with the current upper bounds on $G\mu$ (see equation (1.1)). Using $\alpha = 0.1\gamma G\mu$ only weakens the bound on $G\mu$ by $\sim 20\%$ due to the strong power-law dependence on $G\mu$. Similarily, other uncertainties in the analytic model for the evolution of the cosmic string network used in [7] will only slightly weaken the bound on $G\mu$. Note that we have not included any of the uncertainties in the observational data, or in the analysis of MacGibbon and Carr [12], which will slightly increase the uncertainty in the limits on $f_{\text{BH, obs}}$ and $G\mu$.

## VI. CONCLUSION

We have estimated the fraction of cosmic string loops which collapse to form black holes as a function of their mass-per-unit length, $\mu$, by numerically investigating $25,576$ realistic loop trajectories. The functional form of this fraction has been found, and is well approximated by a power law. Combining this power law with the bound on the fraction of cosmic string loops allowed to form black holes due to the observation of cosmic rays gives a bound on the mass-per-unit length of $G\mu \leq 3.1(\pm 0.7) \times 10^{-6}$. This is consistent with the current bounds on $G\mu$ due to the nucleosynthesis, pulsar timing, and microwave background anisotropy constraints.

Finally, we note that there are several mechanisms for the formation of black holes from cosmic string loops which we have not considered. First, we have not considered the possibility that a black hole may form which encloses only a part of the string loop [13]. Second, we have not included the possibility that a string may "snap", producing a pair of black holes at the string terminals. However, the rate of production of black holes through these mechanisms is estimated to be negligibly low [14]. A third consideration is the long-term effect of gravitational back-reaction on a cosmic string loop's evolution. By smoothing out the small scale structure on the loop, the gravitational back-reaction may enhance the likelihood that the loop will collapse into a small region. Depending on the effectiveness of this process, this may slightly increase the over all fraction of cosmic string loops which collapse to form black holes.

## ACKNOWLEDGMENTS

We would like to thank Bruce Allen for many useful conversations. The work of RRC was supported by PPARC through grant number GR/H71550. The work of PC was supported in part by NSF Grants No. PHY-91-05935 and PHY-95-07740, and by a grant from the Wisconsin Space Grant consortium and the National Space Grant College Fellowship Program.

FIGURES

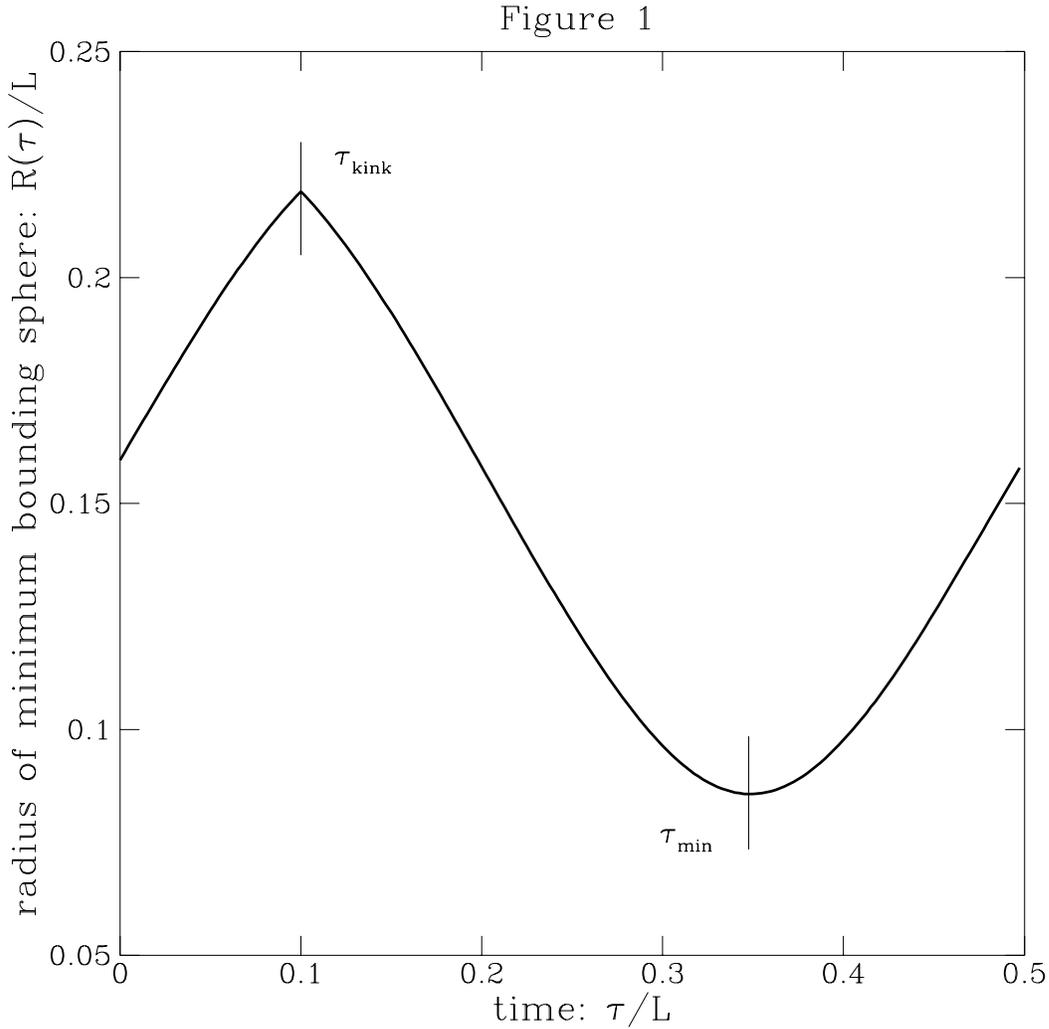

Figure 1

FIG. 1. The radius of the minimum bounding sphere $R(\tau)$ as a function of time $\tau$ for a typical loop during one period of oscillation. For a loop of invariant length $L$, the period of oscillation is $L/2$. All loops formed by loop fragmentation will have at least one left and one right moving kink. The discontinuity in the slope of $R(\tau)$ at time $\tau_{\rm kink}$ corresponds to the instant during the loop's oscillation when a left and right moving kink pass through each other. The loop reaches its minimum radius configuration at time $\tau_{\rm min}$. The radius $R(\tau_{\rm min})$ is used to define the minimum mass-per-unit length, $\mu_{\rm min}$, necessary for the loop to collapse to form a black hole.

Figure 2

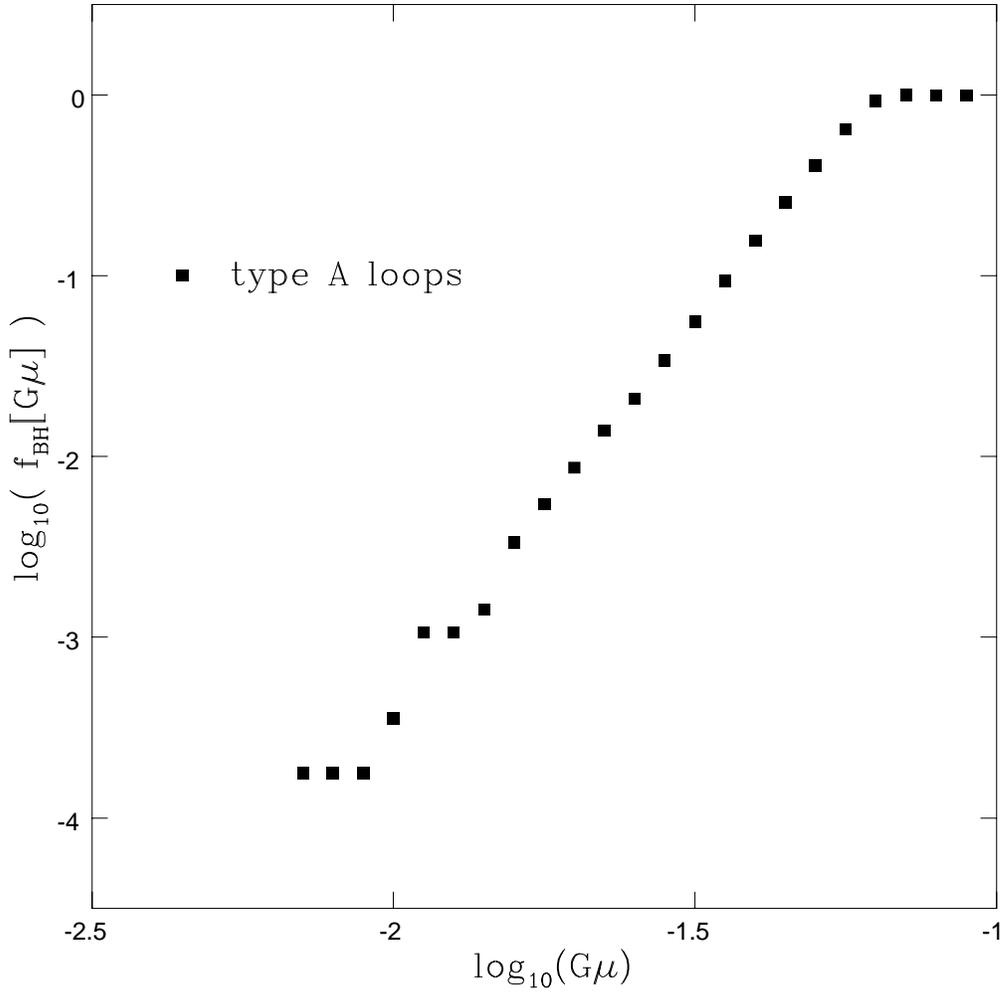

FIG. 2. A log − log plot showing the fraction of the 5,654 child loops, descended from type A parents, with $\mu_{\min} \leq \mu$ for a given value of $\mu$.

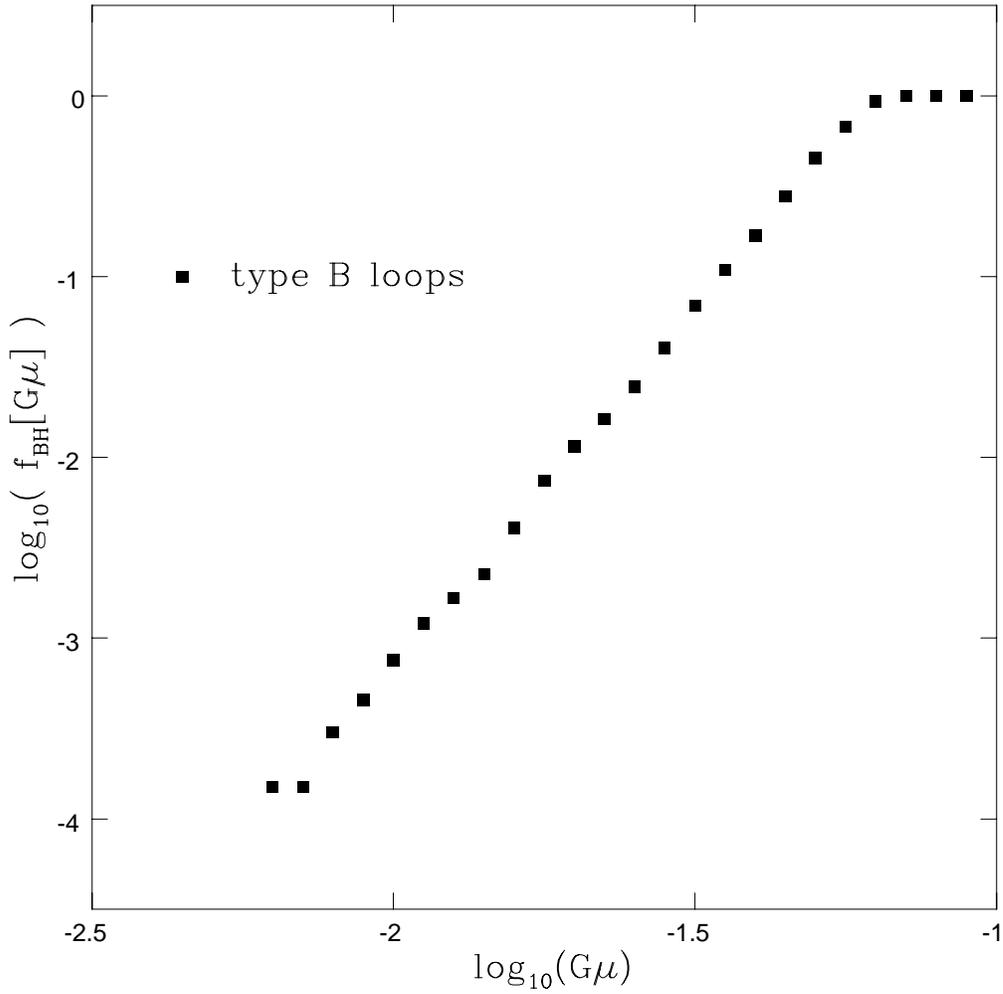

FIG. 3. A log − log plot showing the fraction of the 6,622 child loops, descended from type B parents, with $\mu_{\min} \leq \mu$ for a given value of $\mu$.

Figure 4

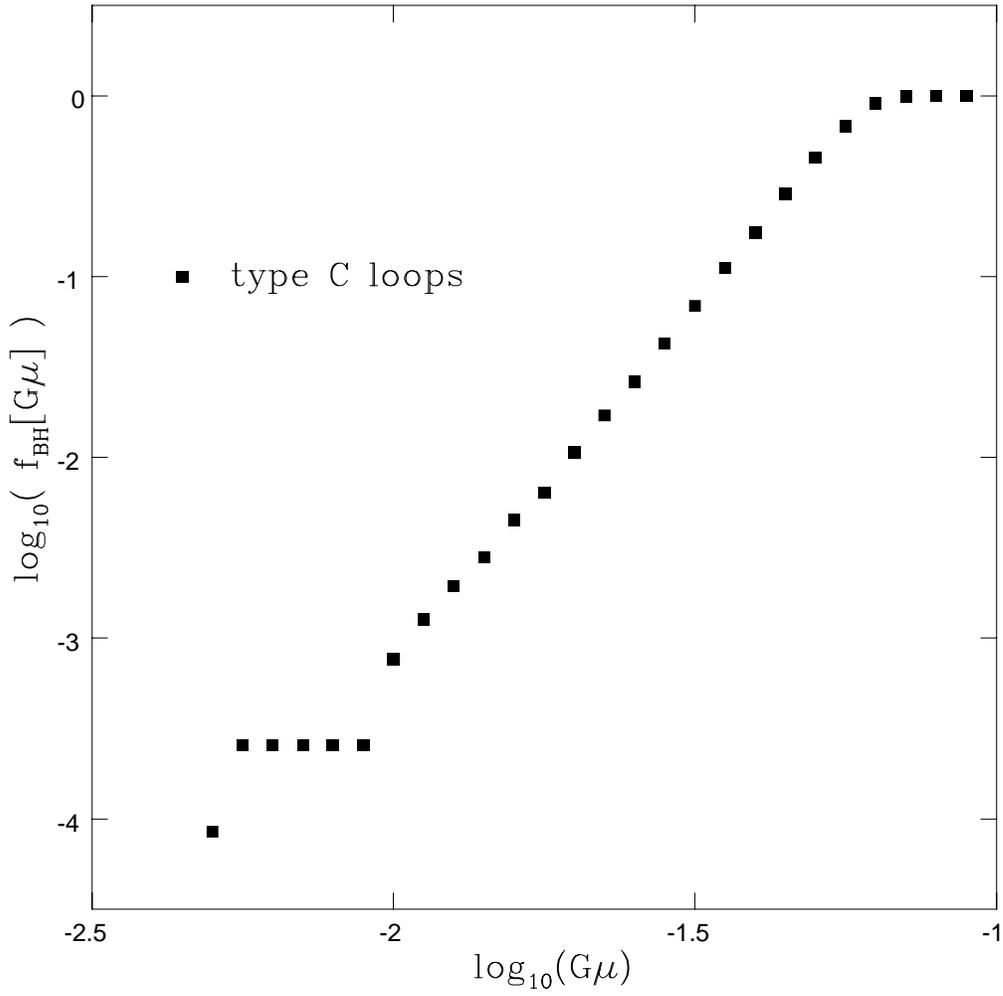

FIG. 4. A log − log plot showing the fraction of the 11,779 child loops, descended from type C parents, with $\mu_{\min} \leq \mu$ for a given value of $\mu$.

Figure 5

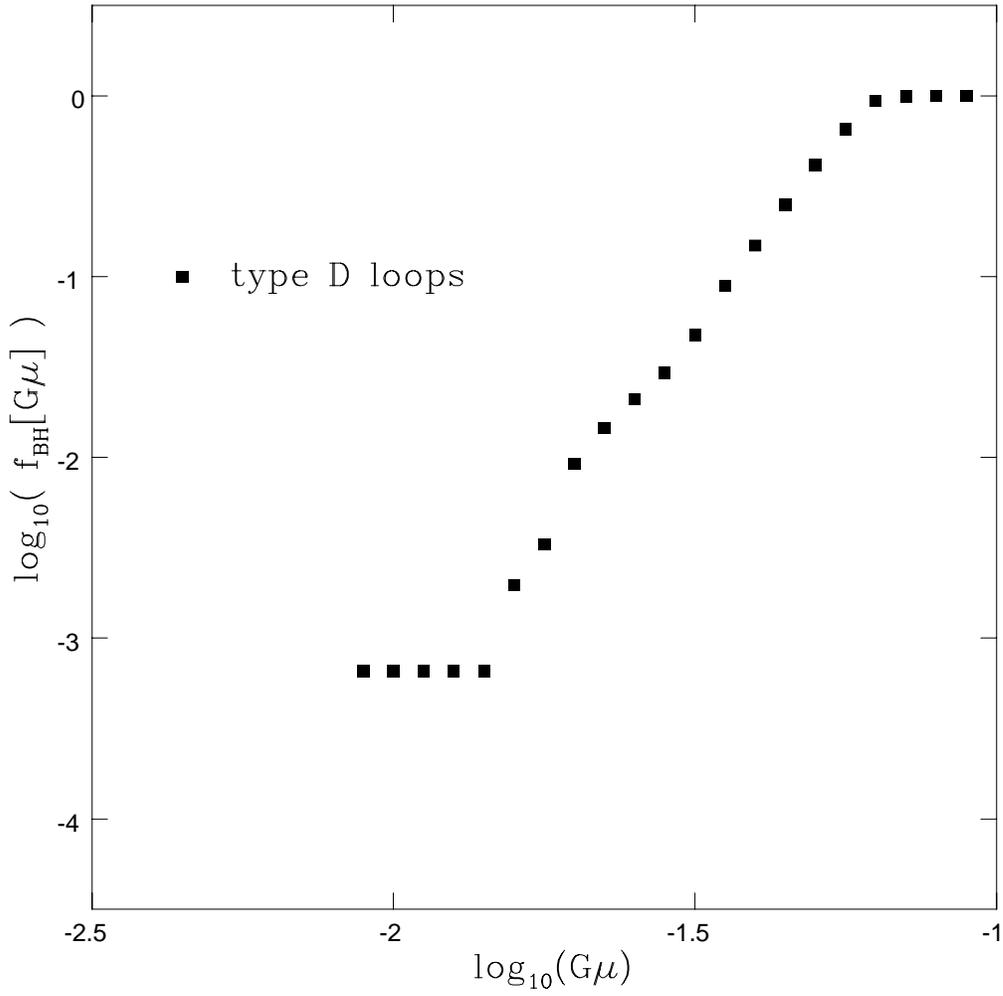

FIG. 5. A log − log plot showing the fraction of the 1,521 child loops, descended from type D parents, with $\mu_{\min} \leq \mu$ for a given value of $\mu$.

Figure 6

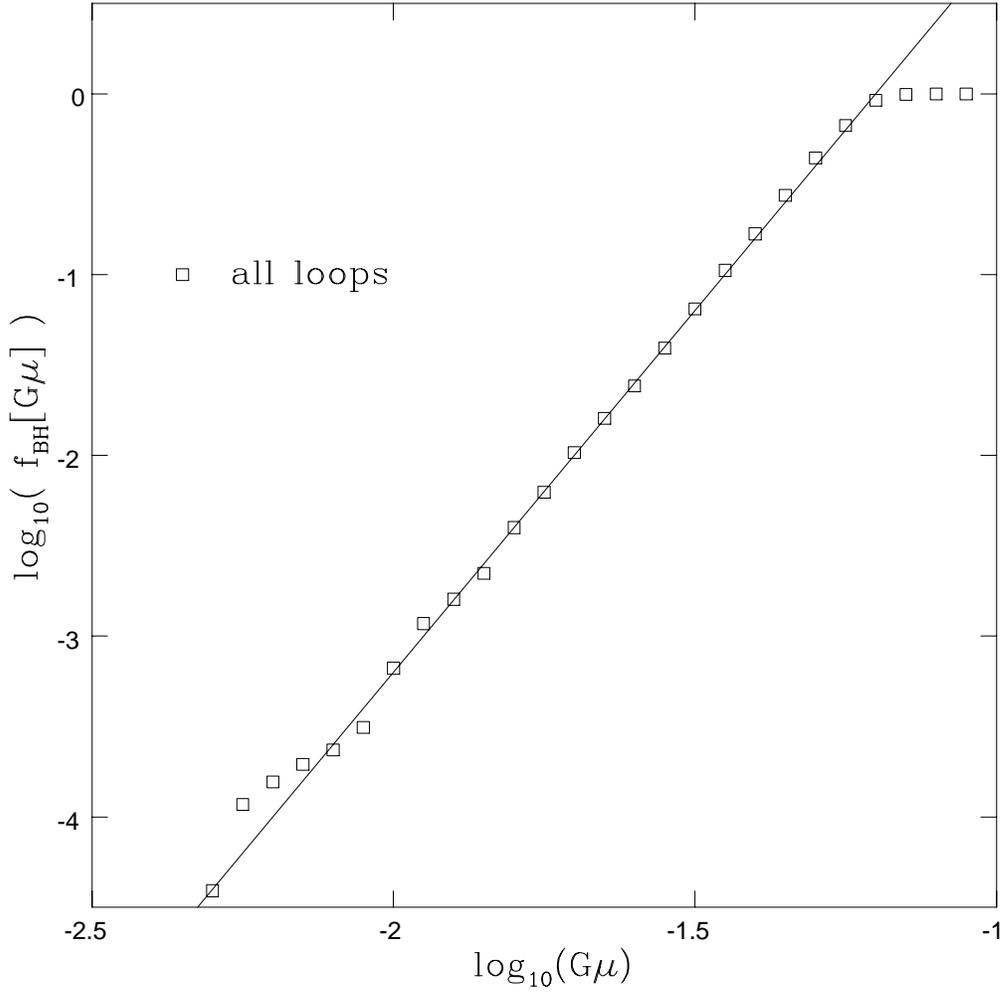

FIG. 6. A log − log plot showing the fraction of the 25,576 child loops, descended from all four parent types, with $\mu_{\min} \leq \mu$ for a given value of $\mu$. The best-fit, $f_{\rm BH} = 10^{4.8}(G\mu)^{4.0}$, to the power-law region of the curve is also shown.

Figure 7

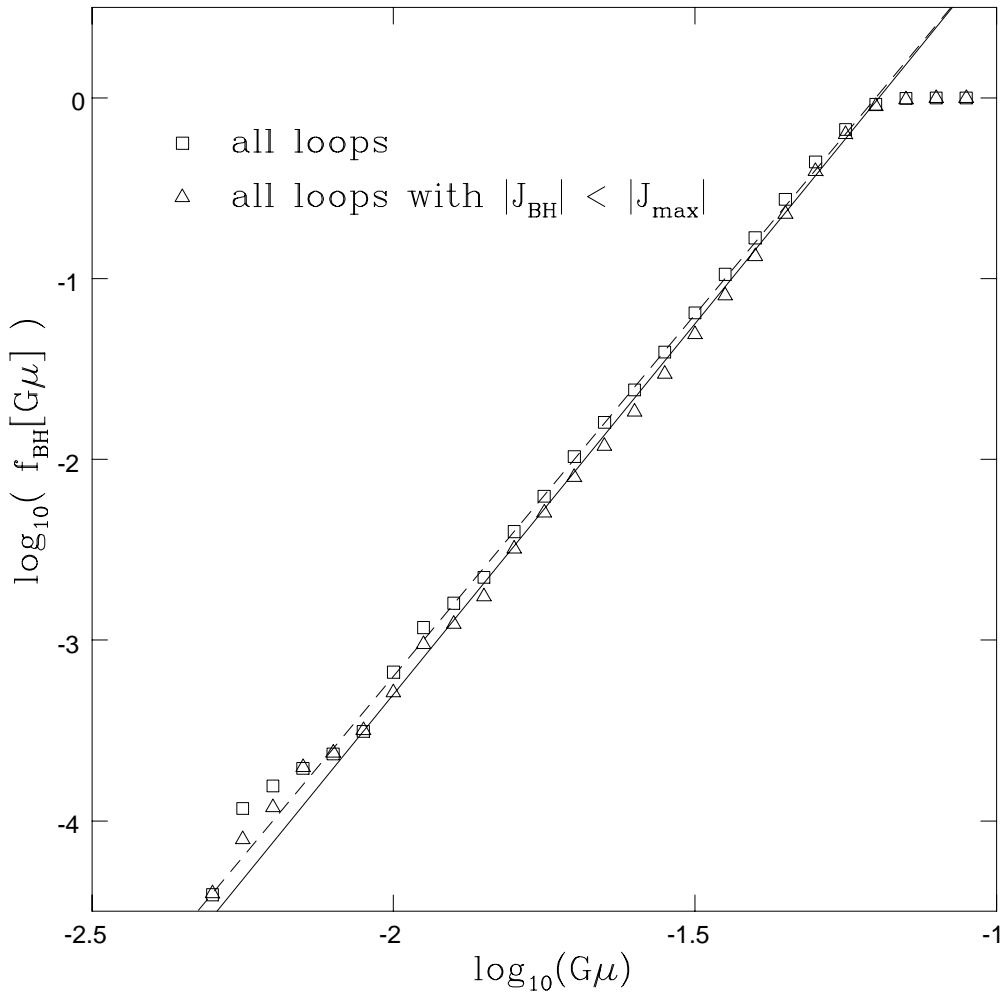

FIG. 7. Plots showing the fraction of the 25,576 child loops, descended from all four parent types, with $\mu_{\min} \leq \mu$ both before (open squares) and after (open triangles) the angular momentum constraint has been applied. The curve $f_{\rm BH}(G\mu) = 10^{4.9}(G\mu)^{4.1}$, used to extrapolate the fraction down to $G\mu = 10^{-6}$ is given by the solid line. The best-fit curve from figure 6, without the angular momentum constraint, is given by the dashed line.